\documentclass[aps,prb,twocolumn,showpacs]{revtex4}
\usepackage{epsfig}
\begin{document} 
\title{Spin-charge coupling in a band ferromagnet: \\ 
magnon-energy reduction, anomalous softening, and damping}
\author{Sudhakar Pandey}
\email{spandey@iitk.ac.in} 
\author{Avinash Singh}
\affiliation{Department of Physics, Indian Institute of Technology Kanpur - 208016}
\begin{abstract}
The effects of correlation-induced coupling between spin and charge fluctuations on spin-wave excitations in a band ferromagnet are investigated by including self-energy and vertex corrections within a systematic inverse-degeneracy expansion scheme which explicitly preserves the Goldstone mode. Arising from the scattering of a magnon into intermediate 
spin-excitation states (including both magnon and Stoner excitations) accompanied with charge fluctuations in the majority spin band, this spin-charge coupling results not only in a substantial reduction of magnon energies but also in anomalous softening and significant magnon damping for zone-boundary modes lying within the Stoner gap. Our results are in good qualitative agreement with recent spin-wave excitation measurements in colossal magneto-resistive manganites and ferromagnetic ultrathin films of transition metals. 
\end{abstract}
\maketitle
\section{Introduction}
Recent observations\cite{Anil-PRL03,Anil-PRB05,Kirschner-PRL07} of large wave-vector spin-wave (magnon) excitations in ferromagnetic ultrathin films of transition metals using spin-polarized electron energy loss spectroscopy (SPEELS) are of crucial importance from many perspectives. For example, apart from providing insight into the microscopic mechanism of ferromagnetic ordering, which can be of direct relevance in context of recent interest in ferromagnetic nanostructures having potential technological applications for magneto-electronic devices,\cite{Book-Spin-Dynamics-06} these observations are also of fundamental importance in understanding the electron spin dynamics in itinerant ferromagnets.\cite{Book-Moriya} 
This is because these large wave-vector excitations, which distinguish an itinerant ferromagnet from the relatively well understood insulating (Heisenberg) ferromagnet, have remained experimentally unexplored in the past owing to certain characteristic features such as heavy damping and large excitation energy.

Theoretical investigations of spin dynamics in these ultrathin films have been carried out mostly by considering transverse spin fluctuations at the level of random phase approximation (RPA) in the ferromagnetic state of the Hubbard model.\cite{Mathon-PRB01,Mills-PRB02,Mills-PRB03-06} However, due to neglect of the strong correlation effects in itinerant ferromagnets, RPA is well known to over-estimate the spin-wave energy, spin-stiffness, and Curie temperature etc., as explicitly demonstrated in recent theoretical investigations by incorporating correlation effects beyond RPA.\cite {AS-PRB06,SP1-PRB07} Indeed, signature of inherent many-body effects have  been found in recent SPEELS\cite{Kirschner-PRL07} and angle-resolved photoemission spectroscopy \cite{Schafer-PRL04,Schafer-PRB05,Cui-Surface-07}
(ARPES) studies in the ferromagnetic phase of Fe. These experimental findings of the signature of strong correlation effects, for example, the much lower magnon energy in Fe film than predicted theoretically at RPA level, as observed in SPEELS,\cite{Kirschner-PRL07} and the quasiparticle mass enhancement and reduced bandwidth in comparison to that predicted within the density functional theory (DFT), as observed in ARPES studies,\cite{Schafer-PRB05} have provided substantial indication of the electron-magnon coupling as the possible origin.

Spin dynamics in the metallic ferromagnetic phase of colossal magnetoresistive (CMR) manganites has also attracted considerable current interest.\cite{Zhang-JPCM-07} Recent spin-wave excitation measurements have revealed several anomalous features in the magnon spectrum near the Brillouin-zone boundary.\cite{Hwang-PRL98,Dai-PRB00,Tapan-PRB02,Ye-PRL06,Ye-PRB07,Moussa-PRB07} 
These observations are of the crucial importance for a quantitative understanding of the carrier-induced spin-spin interactions and magnon damping, and have highlighted the possible limitations of various existing theoretical approaches. 
For example, the prediction of magnon-phonon coupling as the origin of magnon damping\cite{Dai-PRB00} 
and disorder as the origin of zone-boundary anomalous softening\cite{Furukawa-PRB05} 
have been ruled out in recent experiments.\cite{Ye-PRL06,Ye-PRB07,Moussa-PRB07}
Furthermore, the dramatic difference in the sensitivity of long-wavelength and
zone-boundary magnon modes on the density of mobile charge carriers has emerged as one of
the most puzzling feature. Observed for a finite range of carrier concentrations, while the
spin stiffness remains almost constant, the softening and broadening of the 
zone-boundary modes show substantial enhancement with increasing hole concentration.\cite{Ye-PRL06,Ye-PRB07}

Most of the theoretical investigations of spin dynamics in these ferromagnetic manganites have been carried out in the strong coupling (double-exchange) limit ($J/W \gg 1$) of the ferromagnetic Kondo lattice model (FKLM), where mobile ($e_g$) electrons in a partially filled band (of bandwidth $W$ ) are coupled ferromagnetically (with exchange interaction $J$) to the localized core ($t_{2g}$) spins, using a variety of approaches.\cite{Kapetanakis-PRB06} Although providing a good description of magnon damping, these investigations, however, could not satisfactorily account for the observed zone-boundary anomalous softening. In a recent variational investigation, anomalous softening has been demonstrated to be pronounced only in the intermediate-coupling regime ($J/W \sim 1$).\cite{Kapetanakis-PRB06} Furthermore, in this intermediate coupling regime, by taking into account the Coulomb repulsion between the mobile electrons, which is the largest energy scale in manganites and often omitted in the conventional FKLM investigations, recent theoretical investigations have also demonstrated the appearance of several realistic features, such as doping dependent asymmetry of the ferromagnetic phase and enhanced zone-boundary anomalous softening, thereby highlighting the importance of correlated motion of electrons on spin dynamics.\cite{Golosov-PRB05,Kapetanakis-PRB07}  

It is therefore of interest to investigate theoretically the influence of correlated motion of charge carriers in a band ferromagnet on the spin-wave excitation spectrum, particularly the short-wavelength modes. 
The objective of the present paper is to investigate the correlation-induced renormalization of spin-wave excitation spectrum over the entire Brillouin zone in the ferromagnetic state of the Hubbard model.
We will incorporate correlation effects in terms of self-energy and vertex corrections 
within a systematic inverse-degeneracy expansion scheme wherein the spin-rotational symmetry and hence the 
Goldstone mode are preserved explicitly.

The Goldstone-mode-preserving approach, discussed earlier in detail,\cite{AS-PRB06}  
is based on the systematic diagrammatic expansion: 
\begin{equation} 
\phi({\bf q},\omega) = \phi^{(0)} + \phi^{(1)} + \phi^{(2)} + ...
\label{eq:expansion}
\end{equation}
for the irreducible particle-hole propagator $\phi({\bf q},\omega)$ using inverse degeneracy ($1/{\cal N}$) as the expansion parameter. The transverse spin fluctuation propagator is then given by:
\begin{equation}
\chi^{-+}({\bf q},\omega) = \frac{\phi({\bf q},\omega)}{1-U\phi({\bf q},\omega)} \; ,
\label{eq:propagator}
\end{equation}
which characterizes the spin excitations in the spontaneously-broken-symmetric state, including both the low-energy (collective) spin-wave excitations and the high-energy (single-particle) Stoner excitations. 

\begin{figure}
\begin{center}
\vspace*{2mm}
\hspace*{-5mm}
\psfig{figure=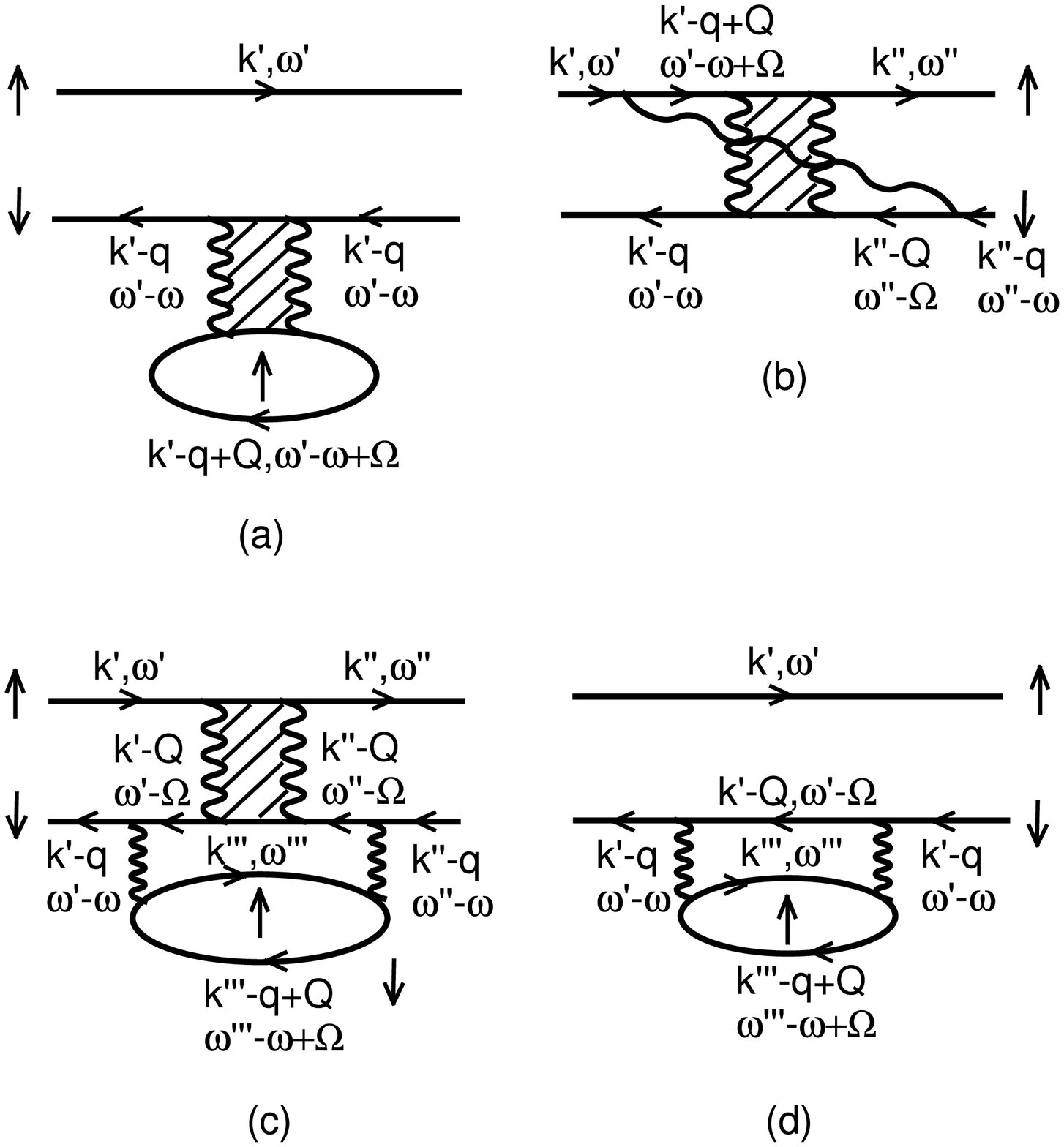,width=90mm}
\vspace*{-15mm}
\end{center}
\caption{First-order quantum corrections to the irreducible particle-hole propagator $\phi({\bf q},\omega)$.}
\label{fig:diagrams}
\end{figure}

In this paper we consider the saturated ferromagnetic state in which the Fermi energy ($\epsilon_{\rm F}$) lies in the majority spin band owing to large exchange splitting such that magnetization $m$ is equal to the particle density $n$. This is similar to the half-metallic ferromagnetic state as observed in the low temperature ferromagnetic phase of many systems such as manganites,\cite{Park-Nature98} and ordered-double perovskites.\cite{Kobayashi-Nature-98} 
In Eq. (1), the bare particle-hole propagator: 
\begin{equation}
\phi^{(0)}({\bf q},\omega) \equiv  \chi^0 ({\bf q},\omega) = \sum_{\bf k}
\frac{1}{\epsilon_{\bf k - q}^{\downarrow +} - \epsilon_{\bf k}^{\uparrow -} + \omega -i \eta} \; , 
\end{equation}
where $\epsilon_{\bf k}^\sigma = \epsilon_{\bf k} - \sigma \Delta$ are the Hartree-Fock (HF) band energies,
$2\Delta= mU$ is the exchange band splitting, and the superscript $+(-)$ refers to particle (hole) states above (below) the Fermi energy $\epsilon_{\rm F}$. 
In terms of this bare particle-hole propagator, the RPA ladder sum: 
\begin{eqnarray}
\chi^{-+}_{\rm RPA} ({\bf q},\omega) &=& \frac{\chi^0({\bf q}, \omega)}{1-U \chi^0({\bf q},\omega)} \nonumber \\
&\approx & \frac{m_{\bf q}}{\omega+\omega_{\bf q}^0 -i\eta} + S({\bf q},\omega) 
\end{eqnarray}
provides a classical (unrenormalized) description of non-interacting spin-fluctuation modes, which include both the low-energy magnon excitations (amplitude $m_{\bf q} \approx m$, energy $\omega_{\bf q}^0$) and the high-energy Stoner excitations $S({\bf q},\omega)$.

The first-order quantum corrections $\phi^{(1)}$, obtained recently for the saturated ferromagnetic state,\cite{AS-PRB06} consist of four distinct processes involving self-energy and vertex corrections, as shown diagrammatically in Fig. 1. The expressions for these diagrams have been given earlier.\cite{AS-PRB06} As required from the spin-rotation symmetry, the net quantum correction $\phi^{(1)}$ vanishes identically for $q,\omega=0$ due to an exact cancellation in order to explicitly preserve the Goldstone mode in the ferromagnetic state. Moreover, this cancellation holds for all $\omega$, indicating no spin-wave amplitude renormalization, as expected for the saturated ferromagnet in which there are no quantum corrections to magnetization.

\section{Spin-charge coupling}
Keeping terms upto first order in $\phi$, the spin-fluctuation propagator (2) can be expressed as:
\begin{equation}
\chi^{-+}({\bf q},\omega) = \frac{1}
{[\chi^{-+}_{\rm RPA} ({\bf q},\omega)]^{-1} - \Sigma({\bf q},\omega)}
\end{equation}
in terms of the first-order magnon self energy $\Sigma({\bf q},\omega) = U^2 \phi^{(1)}({\bf q},\omega)$. From the expressions for the different contributions to the quantum correction $\phi^{(1)}({\bf q},\omega)$, it is seen that the first-order magnon self energy has the following approximate structure:
\begin{equation}
\Sigma({\bf q},\omega) = \sum_{\bf k,Q} \int \frac{d\Omega}{2\pi i} \;
\chi^{-+}_{\rm RPA} ({\bf Q},\Omega)  \Gamma^2 \Pi^0 ({\bf k},{\bf q-Q},\omega-\Omega) \; ,
\end{equation}
highlighting the spin-charge coupling in the ferromagnetic state with the charge fluctuation term:
\begin{equation}
\Pi^0  ({\bf k};{\bf q-Q},\omega-\Omega) = \frac{1}
{\epsilon_{\bf k-q+Q}^{\uparrow +} - \epsilon_{\bf k}^{\uparrow -} + \omega - \Omega - i \eta } 
\end{equation}
in the majority-spin band.
This correlation-induced coupling between the spin and charge fluctuations arises from the scattering of a magnon (with energy $-\omega=\omega_{\bf q}^0$) into intermediate spin-excitation states accompanied by charge fluctuations in the majority spin band. These intermediate states include both the magnon excitations (with energy $-\Omega=\Omega_{\bf q}^0$) and Stoner excitations (spread over the Stoner continuum). This spin-charge coupling is similar to the three body correlations between the Fermi-sea electron-hole pair and a magnon considered in the recent variational investigation.\cite{Kapetanakis-PRB06} 

In Eq. (6), $\Gamma$ represents the interaction vertex for the spin-charge coupling, and is given by: 
\begin{equation}
\Gamma({\bf k};{\bf q},\omega;{\bf Q},\Omega) = U^2 \left ( \chi^0({\bf k};{\bf q},\omega) - \frac{1}{2\Delta '({\bf q},\omega;{\bf Q},\Omega)} \right ) \; ,
\end{equation}
where
\begin{equation}
\frac{1}{2\Delta '({\bf q},\omega;{\bf Q},\Omega)} \equiv \frac{1}{\chi^0 ({\bf Q},\Omega)}
\sum_{\bf k'} \chi^0({\bf k'; q},\omega) \chi^0({\bf k'; Q},\Omega)\; ,
\end{equation}
and
\begin{equation}
\chi^0({\bf k};{\bf q},\omega) \equiv \frac{1}
{\epsilon_{\bf k-q}^{\downarrow +} - \epsilon_{\bf k}^{\uparrow -} + \omega - i\eta} \; .
\end{equation}
This representation of the magnon self energy, with the structure of the spin-charge interaction vertex as in Eq. (8), brings out the similarity with the corresponding result for the ferromagnetic Kondo lattice model,\cite{qfklm} where the term $1/2\Delta '({\bf q},\omega;{\bf Q},\Omega)$ is simply equal to $1/(2\Delta + \omega)$. For the Hubbard model as well, the two terms in the ${\bf k'}$ summation in Eq. (9) decouple for $q=0$, and the term $1/2\Delta ' = 1/(2\Delta + \omega)$. Generally, the term $1/2\Delta '$ has weak momentum dependence due to the averaging over momentum ${\bf k'}$.  

For $q=0$, the spin-charge interaction vertex $\Gamma$ and the magnon self energy vanish identically, and the Goldstone mode is therefore explicitly preserved. For small $q$, $\Gamma^2 \sim ({\bf q}.{\mbox{\boldmath $\nabla$}} \epsilon_{\bf k})^2$, indicating short-range interaction. Also, the spin-charge coupling results in a quantum correction only to the exchange contribution to the spin stiffness as required; quantum corrections to the delocalization contribution of the type $({\bf q}.{\mbox{\boldmath $\nabla$}})^2 \epsilon_{\bf k}$ cancel exactly.\cite{AS-PRB06}

The overall strength of this spin-charge interaction vertex in Eq. (8) is enhanced as $\sim (1/m)^2$ with decreasing band filling $n=m$. This results in an enhancement of the magnon self-energy as $(1/m)^2$, accounting for the two factors of $m$ from the ${\bf k}$ summation and the magnon amplitude in Eq. (6). This behaviour of the magnon self energy with band filling has been investigated quantitatively with respect to both magnon damping and anomalous softening, as discussed later.

To illustrate the correlation-induced renormalization of spin-wave excitations, we have carried out quantitative investigations mostly for the square and simple cubic lattices, with band dispersion:
\begin{equation}
\epsilon_{\bf k} = -2t\sum_\mu \cos (k_\mu a) + 4t'\sum_{\mu < \nu} \cos (k_\mu a) \cos (k_\nu a) \; ,
\label{eq:dispersion}
\end{equation}
yielding respective bandwidths $W=8t$ and $12t$, where $t$ and $t'$ refer to the nearest- and next-nearest-neighbor hoppings, respectively, and $\mu,\nu = x,y,z$. For the spin stiffness calculation, we have also considered the body centered cubic lattice with band dispersion:
\begin{eqnarray}
\epsilon_{\bf k} & = & -8t \cos (k_x a/2) \cos (k_y a/2) \cos (k_z a/2) \nonumber \\
& + & 2t' (\cos k_x a + \cos k_y a + \cos k_z a )\; ,
\end{eqnarray}
and bandwidth $W=16t$. In the following we set $t=1$. Our consideration for $t'$ is motivated by its favorable role in stabilizing the ferromagnetic ordering, as predicted using a variety of approaches.\cite{Vollhardt99-Nolting01}
This has also been demonstrated recently\cite{SP1-PRB07} in the Goldstone-mode-preserving investigation due to reduction in correlation-induced exchange contributions to spin stiffness which have destabilizing tendency on the ferromagnetic state.

\begin{figure}
\begin{center}
\vspace*{-15mm}
\hspace*{-5mm}
\psfig{figure=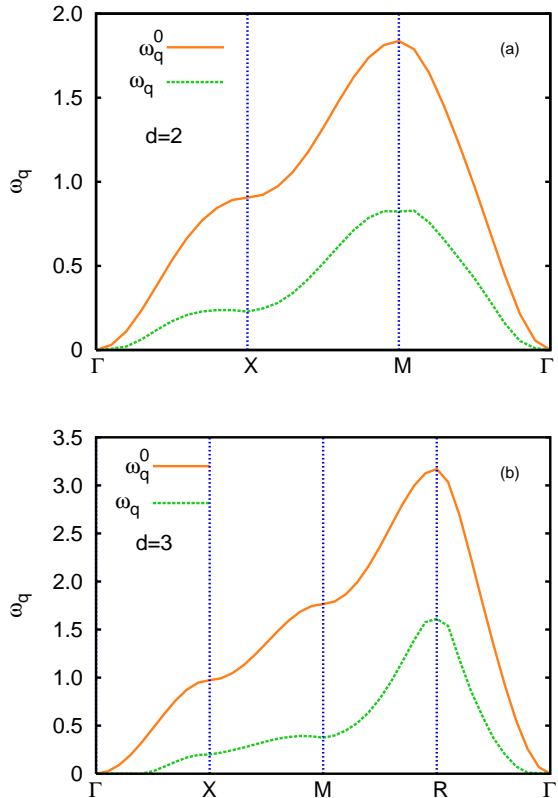,width=90mm}
\vspace*{-8mm}
\end{center}
\caption{Renormalized magnon energy ($\omega_{\bf q}$) is reduced significantly in comparison to the bare (RPA) energy ($\omega_q^0$) due to correlation-induced spin-charge coupling, as shown for square (a) and simple cubic (b) lattices, highlighting the overestimation of RPA, as also reported in a recent spin-wave excitation measurement on the ultrathin film 
of Fe.\cite{Kirschner-PRL07} These results have been obtained for $n=0.5$ and $U/W=1.5$, 
with $t'=0.45$ for the square lattice and $0.25$ for the cubic lattice.}
\label{fig:renormalized-spectrum}
\end{figure}

\section{Renormalized magnon spectrum}
The renormalized magnon energy $\omega_{\bf q}$ for mode $\bf q$ is obtained from the pole condition [$1-U \Re \phi ({\bf q},-\omega_{\bf q})=0$] in Eq. (\ref {eq:propagator}) which also corresponds to the peak in the magnon spectral function 
$A_{\bf q}(\omega)=\frac{1}{\pi} {\rm Im} \chi^{-+} ({\bf q},-\omega)$, the broadening of which provides a quantitative measure of magnon damping, as discussed in section V. The numerical evaluation of the quantum correction $\phi^{(1)}$ by integrating over the intermediate $({\bf Q},\Omega)$ states has been discussed earlier.\cite{SP1-PRB07} 
We note that the evalution of $\phi^{(1)}$ was carried out by including contributions of both the magnon and Stoner excitations.

We find that the renormalized magnon energy $\omega_{\bf q}$ is substantially lower in comparison to the bare (RPA) magnon energy $\omega_{\bf q}^0$ throughout the Brillouin zone, as shown in Fig. \ref{fig:renormalized-spectrum}. This highlights the need to incorporate the strong renormalization due to spin-charge coupling in realistic comparisons. Indeed, in recent SPEELS studies,\cite{Kirschner-PRL07} the measured spin-wave energies in ultrathin films of Fe were found to be significantly smaller than the RPA-level result.

For an orbitally-degenerate ferromagnet (such as Fe with five 3d orbitals per site), the bare and renormalized magnon dispersions $\omega_{\bf q}^0$ and $\omega_{\bf q}$ shown in Fig. 2 provide upper and lower bounds. This is because the first-order quantum correction $\phi^{(1)}$ is suppressed by the factor $1/{\cal N}$ for an ${\cal N}$-orbital-per-site system,\cite{AS-PRB06} and therefore with increasing ${\cal N}$ the renormalized dispersion approaches the bare (classical) dispersion $\omega_{\bf q}^0$ from below as ${\cal N} \rightarrow \infty$. 

\begin{figure}
\begin{center}
\vspace*{-2mm}
\hspace*{0mm}
\psfig{figure=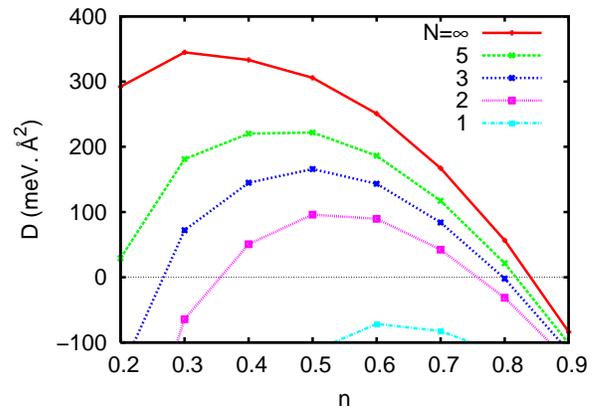,width=80mm}
\vspace*{-5mm}
\end{center}
\caption{Renormalized spin stiffness for different number of orbitals ${\cal N}$, showing the $1/{\cal N}$ suppression of quantum corrections with orbital degeneracy, evaluated for the bcc lattice with bandwidth $W=16t=3.2$eV, Coulomb interaction energy $U=W=3.2$eV, and lattice parameter $a=2.87$\AA\ for Fe. The measured value for Fe is 280 meV.\AA$^2$.}
\end{figure}

The above $1/{\cal N}$ suppression of quantum corrections was obtained for the ${\cal N}$-orbital model with identical intra-orbital interaction ${\bf S}_{i\alpha} . {\bf S}_{i\alpha} $ and inter-orbital interaction (Hund's coupling) ${\bf S}_{i\alpha} . {\bf S}_{i\beta} $.\cite{AS-PRB06} For arbitrary Hund's coupling $J$, the quantum correction factor has been obtained recently,\cite{hunds} and is approximately given by the expression $(U^2 + ({\cal N}-1)J^2)/(U+({\cal N}-1)J)^2$, which rapidly approaches $1/{\cal N}$ with increasing Hund's coupling $J$, particularly for large ${\cal N}$. 

We have quantitatively examined the role of this orbital degeneracy and $1/{\cal N}$ suppression of quantum corrections on the spin stiffness. Fig. 3 shows the renormalized spin stiffness $D=D^{(0)} - \frac{1}{\cal N} D^{(1)}$ evaluated for different number of orbitals ${\cal N}$, where $D^{(0)}$ refers to the bare spin stiffness and $D^{(1)}$ to the first-order quantum correction.\cite{AS-PRB06,SP1-PRB07} Here we have considered a bcc lattice with $t'/t =0.5$, bandwidth $W=16t=3.2$eV, Coulomb interaction energy $U=W=3.2$eV, and the lattice parameter $a=2.87$\AA\ for Fe. These parameter values are close to those considered in a recent investigation of spin-wave excitations in Fe using a realistic band-structure calculation,\cite{naito_2007} where the interaction energy considered is $U=2.13$eV (so that the magnetic moment evaluated per Fe atom is equal to $2.12\mu_{\rm B}$) and the bandwidth from the calculated DOS plot is seen to be about 4eV. Our calculated values for the renormalized spin stiffness for ${\cal N}=5$ are close to the measured value 280meV.\AA$^2$ for Fe.\cite{collins_1969} 

\begin{figure}
\begin{center}
\vspace*{-15mm}
\hspace*{-5mm}
\psfig{figure=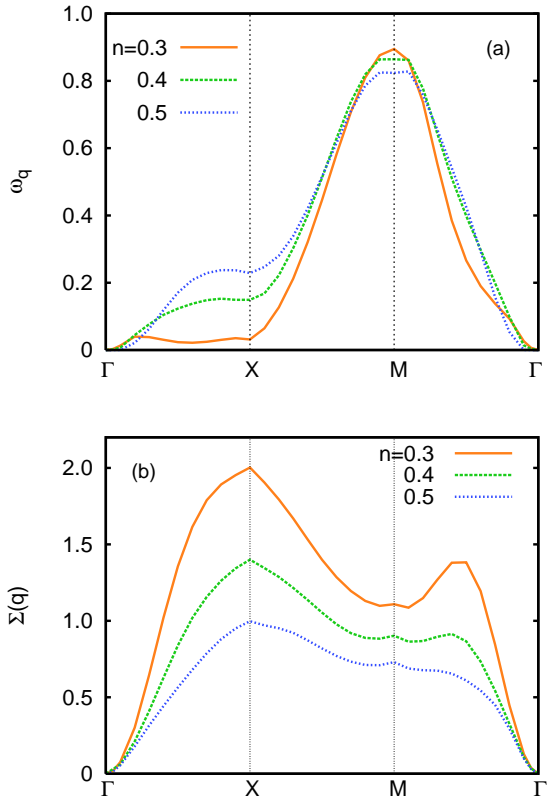,width=90mm}
\vspace*{-8mm}
\end{center}
\caption{Renormalized magnon spectrum for the square lattice (a) shows anomalous softening along the $\Gamma-X$ direction, which becomes more pronounced with decreasing band filling $n$, arising from an anomalous momentum dependence of the magnon self energy (b), which has a maximum at X and similar filling dependence. Here $t'=0.45$ and $U/W=1.0$.} 
\end{figure}

\section{Self-energy correction and anomalous softening}
In addition to magnon-energy reduction due to quantum corrections (Fig. 2), the renormalized magnon spectrum also shows significant anomalous softening near the zone boundary, particularly along the $\Gamma$-X direction, highlighting the anomalous momentum dependence of the quantum correction. While the bare magnon dispersion shows nearly Heisenberg-model behaviour, with magnon energies at X and M (for $d=2$) in the ratio 1:2, and at X, M, and R (for $d=3$) in the ratio 1:2:3, the renormalized magnon dispersion clearly shows strong softening at X relative to M and R. This anomalous softening implies that additional exchange couplings $J_2,J_3,J_4$ etc. must be included in order to describe the magnon dispersion in terms of an effective localized-spin model. Interestingly, in addition to CMR manganites, significant anomalous softening near the zone boundary has also been reported in recent spin-wave dispersion measurement along Fe[001] for Fe film on W(110) by SPEELS.\cite{Kirschner-PRL07}

Now we investigate the effect of carrier concentration on anomalous softening in the context of CMR manganites. We find that the zone boundary anomalous softening along the $\Gamma-X$ direction is enhanced substantially with decreasing band filling, as shown in Fig. 4a. Indeed, such behaviour has been observed in recent spin-wave excitation measurements in the ferromagnetic phase of manganites.\cite{Ye-PRL06,Ye-PRB07} 

This zone-boundary anomalous softening is a direct consequence of the anomalous momentum dependence of the static magnon self energy $\Sigma = U^2 \phi^{(1)}$, as shown in Fig. 4b. The large enhancement in the magnon self energy at X yields a large reduction in the renormalized magnon energy. The substantial enhancement in the anomalous softening with decreasing band filling $n$ is due to $\sim 1/n^2$ enhancement of spin-charge coupling as discussed below Eq. (10). These results for zone-boundary anomalous softening are in agreement with the variational calculation,\cite{Kapetanakis-PRB07} where a proper account of correlated electron motion was found to be necessary for the ferromagnetic manganites. 

\begin{figure}
\begin{center}
\vspace*{-0mm}
\hspace*{-5mm}
\psfig{figure=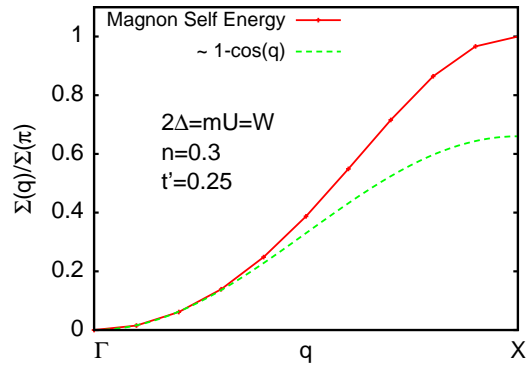,width=70mm}
\vspace*{0mm}
\end{center}
\caption{The magnon self energy $\Sigma({\bf q},\omega=- \omega_{\bf q}^0)$ for the sc lattice shows significant anomalous enhancement near the zone boundary compared to the Heisenberg form, implying anomalous softening of the renormalized magnon energy.}
\end{figure}

Figure 5 shows the magnon self energy $\Sigma({\bf q})$ for the sc lattice. In view of the observed anomalous softening in ferromagnetic manganites, we focus on the $q$-dependence in the $\Gamma - X$ direction and compare with the Heisenberg form $(1-\cos q)$ corresponding to nearest-neighbour coupling. We find that the bare magnon dispersion is nearly Heisenberg like and shows a weak dependence on band filling. However, the magnon self energy shows appreciable enhancement at the zone boundary in comparison with the Heisenberg form, implying zone-boundary softening of the renormalized magnon energy $\omega_{\bf q}^0 - m \Sigma ({\bf q})$. Here the magnon self energy $\Sigma({\bf q},\omega)$ was evaluated at the bare magnon energy $\omega = - \omega_{\bf q}^0$. In contrast, the static self energy evaluated with $\omega=0$ shows no such zone-boundary enhancement, highlighting the role of dynamical effect. Furthermore, we find that this dynamical effect on anomalous softening becomes less pronounced with increasing band filling. 

\section{Correlation-induced magnon damping}
We now turn to the role of spin-charge coupling on magnon damping. At the RPA level, magnon damping is absent for low-energy modes at zero temperature, and arises only at energies above the Stoner gap due to decay into Stoner excitations. However, spin-charge coupling results in finite magnon damping in a band ferromagnet even for magnon modes lying within the Stoner gap. Considering in Eq. (6) only the contribution of collective excitations (4) for simplicity, we obtain the imaginary part of the magnon self energy:
\begin{equation}
\frac{1}{\pi} {\rm Im}\Sigma({\bf q},\omega) = \sum_{\bf k,Q} m_{\bf Q} \Gamma^2 \delta ({\epsilon_{\bf k-q+Q}^{\uparrow +} -
\epsilon_{\bf k}^{\uparrow -} + \omega + \Omega_{\bf Q}^0 }) \; ,
\label{eq:phi-1-magnon}
\end{equation}
which yields finite magnon damping and linewidth, arising from the scattering of a magnon (energy $\omega_{\bf q}^0$)
into intermediate magnon states (energy $\Omega_{\bf Q}^0$) accompanied with charge fluctuations 
(energy $\epsilon_{\bf k-q+Q}^{\uparrow +} - \epsilon_{\bf k}^{\uparrow -}$) in the majority spin band.
Magnon damping is further enhanced when the contribution of Stoner excitations is also included.

\begin{figure}
\begin{center}
\vspace*{-5mm}
\hspace*{-5mm}
\psfig{figure=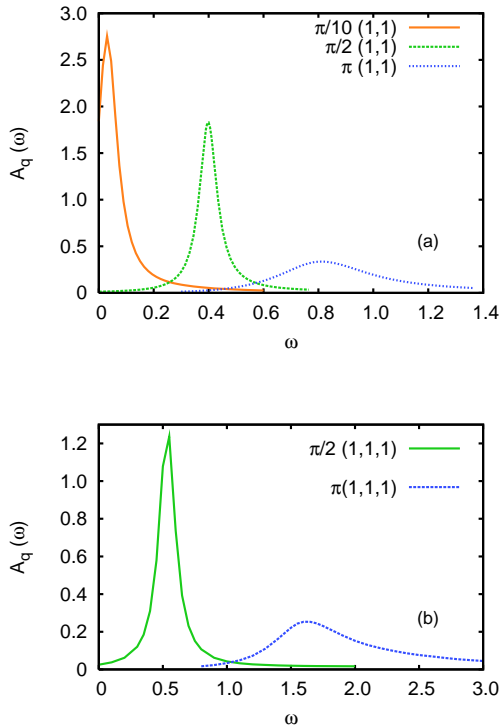,width=80mm}
\vspace*{-8mm}
\end{center}
\caption{Correlation-induced spin-charge coupling results in substantial damping of magnon modes which becomes more pronounced near the zone boundary, shown here for (a) square lattice with $t'=0.45$, $n=0.4$ and $U/W=1.0$, and (b) simple cubic lattice with $t'=0.25$, 
$n=0.5$ and $U/W=1.5$.}  
\end{figure}

We have quantitatively examined magnon damping in terms of the magnon spectral function $A_{\bf q}(\omega) = \frac{1}{\pi} {\rm Im} \chi^{-+} ({\bf q},-\omega)$ using Eq. (2) by including the contribution of both the magnon and Stoner excitations.
In order to highlight the role of correlation-induced magnon damping, we have considered relatively large Stoner gap so that magnon damping is absent at the RPA level. Figure 6 shows the renormalized magnon spectral function $A_{\bf q}(\omega)$ which is substantially broadened near the zone-boundary. This is in broad agreement with the experimental observations of magnon damping in ultrathin transition metal films,\cite{Anil-PRL03, Anil-PRB05,Kirschner-PRL07} and ferromagnetic manganites.\cite{Hwang-PRL98,Dai-PRB00,Tapan-PRB02,Ye-PRB07} 

\begin{figure}
\begin{center}
\vspace*{-15mm}
\hspace*{-5mm}
\psfig{figure=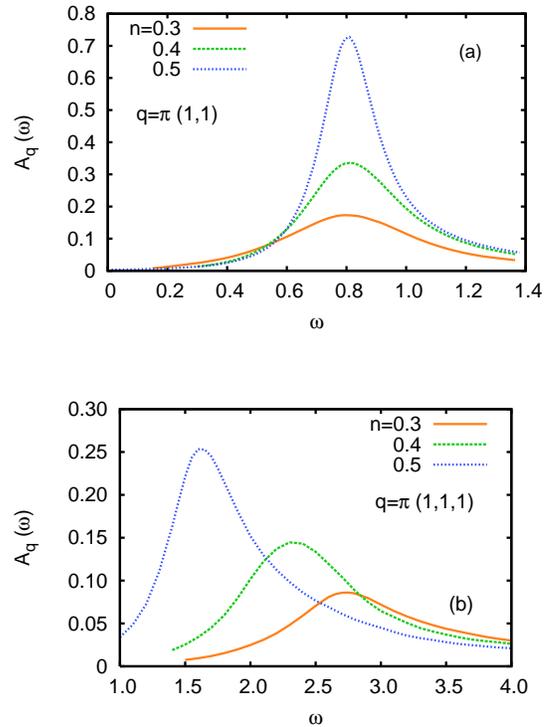,width=84mm}
\vspace*{-8mm}
\end{center}
\caption{Correlation-induced magnon damping becomes more pronounced with decreasing band filling, as shown for the zone-boundary modes for (a) square lattice with $t'=0.45$ and $U/W=1.0$, and (b) simple cubic lattice with $t'=0.25$ and $U/W=1.5$.}   
\label{fig:damp-filling}
\end{figure}

We have also investigated the dependence of magnon damping on band filling $n$ for the zone-boundary mode where damping is most pronounced. With decreasing band filling, damping is enhanced substantially as shown in Fig. \ref{fig:damp-filling}, thereby highlighting the role of charge fluctuations in the magnon-damping mechanism. Similar influence of carrier concentration has also been observed recently in the ferromagnetic manganites.\cite{Ye-PRB07} 

Several realistic features such as multilayers and interfaces (due to nonmagnetic substrate and capping layers)
have also been observed to substantially influence spin-wave excitations 
in ferromagnetic ultrathin films of many transition metals.\cite{Urban-PRL01,Woltersdorf-PRL05,Beaujour-PRB06}
These features have been investigated theoretically by taking into account their effects on electronic structure, although only at the RPA level.\cite{Mills-PRB02,Mills-PRB03-06} Therefore an extension of our investigation with detailed electronic band structure is desirable for a more quantitative comparison.

\section{Conclusion}
In conclusion, we have investigated the effects of correlation-induced spin-charge coupling on the spin-wave excitation spectrum in the ferromagnetic state of the Hubbard model by including self-energy and vertex corrections within a Goldstone-mode-preserving scheme. Arising from the scattering of a magnon into intermediate spin excitation states 
(including both magnon and Stoner excitations) accompanied with charge fluctuations in the majority spin band, the spin-charge coupling results not only in substantial reduction of magnon energies, but also in anomalous softening and damping of magnon modes near the zone boundary lying within the Stoner gap. Both the magnon damping and anomalous zone-boundary softening become more pronounced with decreasing band filling.

Even when the bare magnon dispersion showed nearly Heisenberg form, the renormalized dispersion was shown to exhibit strong softening at X relative to M (for $d=2$) and R (for $d=3$). This anomalous softening at X was shown to be a direct consequence of an anomalous enhancement of the magnon self energy, and implies that the correlated motion of electrons "generates" additional exchange couplings $J_2,J_3,J_4$ etc. within an equivalent localized-spin model with the same magnon dispersion.

The strong $1/{\cal N}$ suppression of quantum corrections due to orbital degeneracy was highlighted by an evaluation of the renormalized spin stiffness for different orbital number ${\cal N}$. For the ${\cal N}=5$ orbital case relevant for Fe, and using realistic bandwidth, Coulomb interaction and lattice parameter values, the quantum correction to spin stiffness was found to be about 25\% at optimal filling. This provides an estimate of the quantum suppression involved in the measured spin stiffness value 280meV.\AA$^2$ of Fe,\cite{collins_1969} arising from the spin-charge coupling. 

These results are of qualitative interest for the ferromagnetic CMR manganites and transition-metal ultrathin films in the context of the observed magnon-energy reduction, anomalous zone-boundary softening, and magnon damping, highlighting the influence of correlated electron motion on their spin dynamics. However, several realistic features need to be incorporated for a  quantitative comparison with experiments. These include, for example, in case of manganites the Hund's coupling of the $e_g$ electrons with the core ($t_{2g}$) spins, and their orbital degree of freedom which was predicted to have a major influence on the anomalous softening when coupled with the lattice degree of freedom.\cite{Khaliullin-PRB00} Similarly, for the ultrathin transition-metal films, a realistic electronic description of the magnetic multilayers as well as of the nonmagnetic substrates and capping layers is necessary in view of the accumulating experimental evidence for their substantial influence on the electron spin dynamics.\cite{Urban-PRL01,Woltersdorf-PRL05,Beaujour-PRB06} 

\section*{Acknowledgments}
One of us (SP) gratefully acknowledges financial support from CSIR.

\end{document}